\documentstyle[preprint,aps]{revtex}
\preprint{{\it Published in Europhysics Lett. \bbox{29}, 515 (1995)}}
\begin{document}
\draft \title{Statistics of Oscillator Strengths in Chaotic Systems}

\author{Nobuhiko Taniguchi, A. V. Andreev and Boris L.
  Altshuler} \address{ Department of Physics, Massachusetts Institute of
  Technology,\\ 77 Massachusetts Avenue, Cambridge, MA 02139\\
 and\\
  NEC Research Institute, 4 Independence Way, New Jersey 08540}
\date{
}

\maketitle
\begin{abstract}
  The statistical description of oscillator strengths for systems like
  hydrogen in a magnetic field is developed by using the supermatrix
  nonlinear $\sigma$-model.  The correlator of oscillator strengths is
  found to have a universal parametric and frequency dependence, and its
  analytical expression is given.  This universal expression applies to
  quantum chaotic systems with the same generality
  as Wigner-Dyson statistics.
\end{abstract}

\pacs{PACS number: 05.45.+b, 32.60.+i}


A conventional way to approach the phenomenon of quantum chaos is to study
spectra of quantum systems~\cite{Berry91}.  These spectra are known to
be well described by Wigner-Dyson statistics of energy
levels~\cite{Mehta}, based on the eigenvalue distribution of random matrix
ensembles.
Usually chaos manifests itself in highly excited states of a
system.
Excitation from a regular state (e.g., the ground state)
to a chaotic state can be achieved by applying an external perturbation.
Experiments of this type provide {\em independently\/} the
transition probabilities from a given initial state to excited
states, as well as the spectrum of the system.
These probabilities for optical excitations are known as
oscillator strengths.  Like the eigenenergies, the oscillator strengths
of highly excited states are random, and should be described
statistically. Along this line, the distribution of the transition
matrix elements was predicted to have the Porter-Thomas
distribution in the chaotic limit~\cite{Alhassid86}.

In this Letter we develop a statistical description of oscillator
strengths based on the Wigner-Dyson hypothesis.  We investigate a
statistical characteristic different from Ref.~\cite{Alhassid86} --- the
correlation function of oscillator strengths at {\em different\/} frequencies
and ``magnetic fields''.
It is emphasized that
we make no further assumptions such as on the form of the matrix elements.
Since the information contained within the oscillator strengths is
independent of the spectral properties, their statistics provide a
characterization of quantum chaos which complements the usual energy level
correlations.
We determine universal parametric correlations of oscillator strengths
which should apply with the same generality as Wigner-Dyson statistics. We
have also discovered an unexpected differential relation between this
parametric correlations and the parametric density of states
correlator of Ref.~\cite{Simons93}.

The results of this Letter can be applied to responses of various impurity
states in semiconductors, or to microwave and optical excitations of
chaotic quantum dots etc.  They can be easily modified to describe a much
broader class of phenomena, e.g., cross sections of
nuclear excitations by neutrons.
However, as a concrete example, we have in mind the hydrogen atom in a
magnetic field, which attracts great interest as a ``real'' quantum system
with a chaotic classical analogue.  The energy spectrum of this system is
known to obey Wigner-Dyson statistics~\cite{Hydrogen}.  In addition, the
distribution of the transition matrix elements is observed to obey the
Porter-Thomas distribution~\cite{Wunner89}.

The oscillator strength per unit energy is conventionally defined as
\begin{equation}
  W(\bar\Omega ,\bar B)=2\bar\Omega \sum_\alpha {\left| {\left\langle
      \Phi_0 \right|\hat{z}\left| \alpha \right\rangle} \right|^2\;\delta
    (E_\alpha (\bar B)-E_0(\bar B)-\bar\Omega )},
\label{osc-def}
\end{equation}
where $B$ denotes the magnetic field.  More generally, $B$ can stand for
any external parameter of the system.
In writing down Eq.~(\ref{osc-def}), we use the transition matrix element
between a certain initial state $\left|\Phi_0\right\rangle$ with the
energy $E_0$ and a highly excited state $\left|\alpha \right\rangle$ with
$E_\alpha$ induced by the light polarized along $z$-axis.
However, it is not necessary to assume this particular form of the
transition matrix elements to proceed our theory.

We have evaluated the correlation function of oscillator strengths
\begin{equation}
  S(\Omega ,B)=\left\langle {W(\bar\Omega,\bar B)W(\bar\Omega+\Omega ,\bar
    B+B)} \right\rangle ,
\label{S-def}
\end{equation}
for small energy separation ($\Omega\ll\bar\Omega$) and at slightly
different ``magnetic fields" ($B\ll\bar B$). Here $\langle\cdots \rangle$
denotes the averaging over a certain interval of the spectrum and/or
of the parameter $\bar B$.
The magnitude of the correlator $S(\Omega ,B)$ is
not universal.  It depends on the initial state $\left|\Phi
_0\right\rangle$, on the type of excitation, and on the direction of the
polarization.  (See Eq.~(\ref{defS0}) below.)  However, the dependence of
$S$ on $\Omega$ and $B$ turns out to be determined by a
universal function $ s(\omega ,x)$,
\begin{equation}
  S(\Omega ,B)=S_0 \,\left[s(\omega ,x)+1\right];\qquad S_0=\langle W\rangle^2,
\label{S-result}
\end{equation}
where we use the rescaled parameters defined by~\cite{Simons93}
\begin{eqnarray}
  \omega & &\equiv\Omega /\Delta\:;\quad\quad\quad\: \Delta =\left\langle
  E_{\alpha +1} -E_\alpha\right\rangle \label{defomega}, \\ x & &\equiv
  B\sqrt {C(0)}\:;\quad\: C(0)=\left\langle {(\partial E_\alpha /
\partial B)^2} \right\rangle/\Delta^2. \label{defx}
\end{eqnarray}
The two terms in Eq.~(\ref{S-result}) correspond to the connected and
disconnected parts of the correlator.
The function $s(\omega ,x)$ is classified according to the Dyson
universality classes: orthogonal $(o)$, unitary $(u)$,
or symplectic $(s)$.  It is convenient to present $s(\omega ,x)$ by
\begin{equation}
  s_e(\omega ,x)=\left( {{{\partial ^2} \over {\partial \omega
        ^2}}+2\zeta_e{\partial \over {\partial x^2}}} \right)\,h_e(\omega
  ,x),
  \label{defs}
\end{equation}
where $e=o$, $u$, $s$; $\zeta_o=\zeta_s=2$, $\zeta_u=1$ and
\begin{equation}
  h_u(\omega ,x)={-1\over 2\pi^2}\int_{-1}^1\!d\lambda\!\int_1^\infty
  \!d\lambda _1{\cos \left[ {\pi \omega (\lambda _1-\lambda )} \right]
    \over (\lambda_1-\lambda)^2} \exp \left[-{{\pi ^2x^2} \over 2}(\lambda
  _1^2-\lambda^2) \right],
\label{h-unitary}
\end{equation}
\begin{eqnarray}
  h_{o,s}(\omega ,x)={-1\over \pi^2}
  \int\!\!\!\int\!\!\!\int\!d\lambda d\lambda _1d\lambda _2\;
&&
{|1-\lambda
      ^2|\;\cos \left[ {\pi \omega (\lambda _1\lambda _2-\lambda )}
    \right] \over
    (\lambda_1^2+\lambda_2^2+\lambda^2-2\lambda_1\lambda_2\lambda-1)^2}
\nonumber\\\times&&
\exp \left[-{{\pi ^2x^2} \over 4}|2\lambda _1^2\lambda _2^2-\lambda
  _1^2-\lambda _2^2-\lambda ^2+1| \right].
\label{h-orthogonal}
\end{eqnarray}
\noindent In Eq.~(\ref{h-orthogonal}), the integral region is defined by
$\lambda\in (-1,1)$, $\lambda_{1,2}\in (1,\infty )$ for the orthogonal
case, or $\lambda\in (1,\infty )$, $\lambda_1\in (-1,1)$, and $\lambda_2\in
(0,1)$ for the symplectic case.

Let us compare $S(\Omega ,B)$ with the more conventional statistical
characteristic of quantum chaotic systems. For a certain value of $B$, the
density of states at energy $E$ can be written as $\nu (E, B)=\sum_\alpha
{\delta (E-E_\alpha (B))}$. The parametric correlator of densities of
states is written in terms of rescaled variables
Eqs.~(\ref{defomega},\ref{defx})
\begin{equation}
  \left\langle {\nu (E,\bar B)\,\nu (E+\Omega , \bar B + B
    )} \right\rangle =
 {1 \over {\Delta ^2}}\left[ {1+k( {\omega , x)}}\right]
\label{defK}
\end{equation}
The explicit forms for $k_o (\omega , 0)$, $k _u (\omega , 0)$ and $k _s
(\omega , 0)$ were obtained by Dyson (see, e.g., Ref.~\cite{Mehta}), and
were recently evaluated at finite $x$~\cite{Simons93} by the supermatrix
method~\cite{Efetov83}. In our notation these results can be written
as\cite{endnote}
\begin{equation}
k(\omega ,x)=\partial^2 h(\omega ,x)/\partial \omega ^2.
\label{formulak}
\end{equation}
Both $ k(\omega ,x)$ and $s(\omega ,x) $ are singular at $ \omega =0$ and
$x=0$. We present in Fig.~1 the ratio
$ r(\omega ,x)=(1+s(\omega ,x))/(1+k(\omega ,x))$ which is not singular.

Comparing $s(\omega ,x)$ of Eq.~(\ref{defs}) with $k(\omega ,x)$ of
Eq.~(\ref{formulak}), we get a differential relation:
\begin{equation}
  {{\partial ^2} \over {\partial \omega ^2}} s(\omega ,x)=\left(
  {{{\partial ^2} \over {\partial \omega ^2}}+2\zeta {\partial \over
      {\partial x^2}}} \right)\,k(\omega ,x).
  \label{relation}
\end{equation}
This relation is not trivial at all~\cite{endnote}, especially when
considered from the semiclassical point of view~\cite{Berry91,Gutzwiller}.
That approach is based on the approximation of the density of states $\nu
(E,B)$ by a certain sum over the classical periodic trajectories.
Although the averaged oscillator strength itself may be expressed by the
summation of the periodic trajectories~\cite{Wilkinson87}, the oscillator
strength correlator apparently requires all {\em closed\/} trajectories to
be taken into account, not just the classical {\em periodic\/} ones.  (See
also Ref.~\cite{Bogomolnyi90} for some
discussions.)
Eqs.~(\ref{defs},\ref{formulak}) for $s(\omega ,x)$ and $k(\omega ,x) $
are valid for arbitrary $\omega $ and $x$, including nonperturbative
regime at $\omega\lesssim 1$. We believe that these formulas as well as
Eq.~(\ref{relation}) have wider region of applicability than the
conventional semiclassical approach.

It is worth mentioning that the definition of $h(\omega ,x)$ should be
interpreted rather formally, and in fact, integral expressions
Eq.~(\ref{h-unitary},\ref{h-orthogonal}) for $h(\omega ,x)$ diverge
logarithmically. Of course, the corresponding integrals for $s(\omega ,x)$
and $k(\omega ,x)$ are converged.  However, we have introduced $h(\omega
,x)$ not only to make formulas for $s(\omega ,x)$ and $k(\omega
,x)$ more compact. Note that the finite part of $h(\omega
,x)-h(0,x)$ is related to the generalized parametric number variance
function $v(\omega ,x)$ by (compare with Ref.~\cite{Goldberg91})
\begin{equation}
 h(\omega ,x)-h(0,x)={1\over 2} \left( v(\omega ,x) -v(0,x)\right),
\end{equation}
where $v(\omega ,x) = [ n(\epsilon+\omega ,x+\bar x)-n(\epsilon
,\bar x)-\omega]^2$ and $n(\epsilon ,x)=\sum_i \theta (\epsilon
-\epsilon_i(x))$.

Now let us sketch the derivation of our main
results~(\ref{S-result}-\ref{h-orthogonal}).  One can write $S(\Omega
,B)$,
\begin{equation}
  S(\Omega ,B)=\int \!\!  {d{\bf r}_1d{\bf r}_2d{\bf r}_3d{\bf r}_4 A({\bf
      r}_1,{\bf r}_2,{\bf r}_3,{\bf r}_4)R({\bf r}_1,{\bf r}_2,{\bf
      r}_3,{\bf r}_4)} ,
\end{equation}
where
\begin{eqnarray}
  &&A ={{4\bar\Omega^2} \over {\pi ^2}}z_1z_2z_3z_4\Phi _0^*({\bf
r}_1)\Phi_0({\bf r}_2)\Phi_0^*({\bf r}_3)\Phi_0({\bf r}_4), \\
&&R=\left\langle {\mbox{Im} G_{E_0+\bar\Omega}^R({\bf r}_1,{\bf r}_2;\bar
  B)\,\mbox{Im} G_{E_0+\bar\Omega+\Omega }^R({\bf r}_3,{\bf r}_4;\bar B+B)}
\right\rangle.
\label{defR}
\end{eqnarray}
and $G_E^{R,A}({\bf r},{\bf r'};B) =
\langle {\bf r} |( {E-H(B)\pm i0})^{-1}| {{\bf r'}}\rangle$ are the retarded
and advanced Green's functions.
The factor $A$ is determined by the initial state wave function $\Phi_0
({\bf r})$, whereas $R$ reflects only the
properties of excited states, which are presumed chaotic.
In contrast to the evaluation of $K(\Omega ,B)$ in Eq.~(\ref{defK}), we
now have to accommodate the dependence on four different positions ${\bf
  r}_1$, ${\bf r}_2$, ${\bf r}_3$,${\bf r}_4$ as depicted in Fig.~2a.

We apply the zero-dimensional supermatrix model~\cite{Efetov83} to
determine $R$:
\begin{eqnarray}
  {2 \over \pi^2\nu^2}R &&({\bf r}_1,{\bf r}_2,{\bf r}_3,{\bf r}_4) =
  f(r_{12})f(r_{34})\,\mbox{Re}\left\langle 1- {Q^{11}_{33}Q^{22}_{33}}
\right\rangle_Q \nonumber\\ &&-f(r_{13})f(r_{24})\,\mbox{Re}\left\langle
{Q^{12}_{34}Q^{21}_{43}} \right\rangle_Q
-f(r_{14})f(r_{32})\,\mbox{Re}\left\langle {Q^{12}_{33}Q^{21}_{33}}
\right\rangle_Q,
\label{RinQmatrix}
\end{eqnarray}
where the dimensionless function $f(r)=\langle\mbox{Im}G^R(r)\rangle
/\langle\mbox{Im}G^R(0)\rangle$ which is not universal~\cite{Prigodin94}.
For example, $f(r)$ becomes $\sin (pr) \exp (-r/2\ell)/pr$ in three
dimensional metallic grain with the Fermi momentum $p$ and the mean free
path $\ell$.
$\left\langle\cdots\right\rangle_Q$ of Eq.~(\ref{RinQmatrix}) denotes the
integration over the saddle point manifold $\left\langle
\cdots\right\rangle_Q\equiv \int {DQ\,\left( \cdots \right)}\,\exp \left(-
(i\pi\omega/4)\mbox{STr}Q\Lambda-(\pi^2x^2/32\zeta_e)\mbox{STr}[Q,\Lambda
]^2\right)$, and the definition of $\mbox{STr}$ as well as $\Lambda$ and
the structure of supermatrices $Q$ can be found in~\cite{Efetov83}.
We can identify physically each term of Eq.~(\ref{RinQmatrix}) as
in Fig.~2(b-d), --- DOS fluctuation-type (Fig.~2b),
Cooperon-type (Fig.~2c) and Diffuson-type (Fig.~2d).
In fact, we get the DOS correlator of Eq.~(\ref{defK}) by integrating
Fig.~2b over ${\bf r}_1$ and ${\bf r}_3$, whereas
integrating Fig.~2c and 2d over ${\bf r}_1$ and ${\bf r}_3$ (Fig.~2c) or
${\bf r}_1$, ${\bf r}_2$ (Fig.~2d) gives another function which was
recently shown to correspond to the dielectric function of periodic
systems with chaotic unit cell~\cite{Taniguchi93d}.
Definite integration over $Q$-matrix leads to
$\mbox{Re}\langle {Q^{11}_{33}Q^{22}_{33}}-1 \rangle_Q = -2 -2\partial^2
h(\omega ,x)/\partial \omega ^2$, $\mbox{Re}\langle
{Q^{12}_{33}Q^{21}_{33}}\rangle_Q = -4\,\partial h(\omega, x)/\partial
x^2$, and the Cooperon-type contribution
$\langle{Q^{12}_{34}Q^{21}_{43}}\rangle_Q$ is equal to the Diffuson-type
$\langle{Q^{12}_{33}Q^{21}_{33}}\rangle_Q$ for the orthogonal and
symplectic cases, or vanishes for the unitary.  Everything combined, we
obtain our results (\ref{S-result}-\ref{h-orthogonal}), with the prefactor
$ S_0 $ equal to
\begin{equation}
  S_0 =\left[ {2\bar\Omega\nu\int {d{\bf r}_1d{\bf r}_2z_1z_2\Phi
      _0^*({\bf r}_1)f(r_{12})\Phi _0({\bf r}_2)}} \right]^2,\label{defS0}
\end{equation}
since $\Phi_0({\bf r})$ is real for orthogonal and symplectic cases.
As it was mentioned, the constant $S_0$ as well as the parameters $\Delta$
and $C(0)$ is not universal.  However, the dependence of $S$ on the
rescaled frequency $\omega$ and ``magnetic field'' $x$ is universal.

In conclusion, we have developed the statistical description of oscillator
strengths of a chaotic system such as hydrogen in a magnetic field.  In
particular, we have obtained the analytical form of the correlator of
oscillator strengths $S(\Omega ,B)$, dependent on $\Omega$ and $B$
(Eqs.~(\ref{S-result}-\ref{defs})).
We have found an identity Eq.~(\ref{relation}) that relates $S$ with the
density of states correlator. It looks like a challenge
for the semiclassical theory of quantum chaos to derive this differential
relation.
The universality of $S(\Omega ,B)$ is as general as that of the
correlation function of densities of states $K(\Omega ,B)$, and should be
applicable to various physical situations: impurity states in
semiconductors, to microwave and optical excitations in chaotic quantum
dots, etc.
We believe that experimental studies of the statistics of the oscillator
strength provides a new way of characterizing the chaotic systems,
complimentary to the spectral statistics.

The authors are especially grateful to B. D. Simons for valuable
discussions.  We also like to thank M.  Courtney, D.  Kleppner, E.
Mucciolo, and V. N. Prigodin for helpful discussions.  The work was
partially supported by Joint Services Electronic Program Contract No.
DAAL 03-89-0001.  N.T. also acknowledges partial supprot from the research
fellowship from Murata Overseas Scholarship Foundation.

%
%


\begin{references}

\bibitem{Berry91} M. Berry, in {\em Chaos and Quantum Physics}, edited by
  M.-J. Giannoni, A.  Voros, and J. Zinn-Justin (North-Holland, Amsterdam,
  1991), pp.\ 251.

\bibitem{Mehta} M.~L. Mehta, {\em Random Matrices 2nd ed.} (Academic
    Press, San Diego, 1991).

\bibitem{Alhassid86}
Y.~Alhassid, and R.~D.~Levine, Phys. Rev. Lett. {\bf 57}, 2879 (1986).


\bibitem{Simons93} B.~D. Simons and B.~L. Altshuler, Phys. Rev. B {\bf
    48}, 5422 (1993).

\bibitem{Hydrogen}
D. Wintgen and H. Friedrich, Phys. Rev. Lett. {\bf 57}, 571 (1986);
D. Delande and J. C. Gray, {\it ibid.}, 2006;
G. Wunner {\it et. al.}, {\it ibid}, 3261.
%
For a review with recent results, see D.~Delande, in {\em Fundamental
  Systems in Quantum Optics}, edited by J.~Dalibard, J.~M. Raimond, and
J.~Zinn-Justin (North-Holland, Amsterdam, 1992), pp.\ 381; D.~Kleppner,
{\it ibid.}, pp.\ 417.


\bibitem{Wunner89}
G.~Wunner, W.~Schweizer, and H.~Ruder, in {\em The
    Hydrogen Atom\/} (Springer-Verlag, Berlin, 1989), pp.\ 300.

\bibitem{Efetov83} K.~B. Efetov, Adv. Phys. {\bf 32}, 53 (1983). See also
J.~J.~M. Verbaarschot, H.~A. Weidenm\"{u}ller,
  and M.~R. Zirnbauer, Phys. Rep.  {\bf 129}, 367 (1985).

\bibitem{endnote} As was shown by B. D. Simons {\it et.al.\/}, Nucl. Phys.
  {\bf B409}, 487 (1993), $k(\omega ,x)=\partial^2 h/\partial \omega^2$
  gives the density-density correlator for the one-dimensional gas with
  $1/r^2$ interaction when $\omega\to r$ and
  $-ix^2/2\to t$.  In the same sense, after the Fourier transformation in
  $r$ and $t$, $\partial h/\partial x^2$ is found to give the dynamical
  conductivity $\sigma (q,\omega )$ of this one-dimensional gas with the
  help of the continuity relation.

\bibitem{Gutzwiller} M.~C. Gutzwiller, {\em Chaos in Classical
    and Quantum Mechanics\/} (Springer-Verlag, New York, 1990).

\bibitem{Wilkinson87} M.  Wilkinson, J. Phys. A {\bf 20}, 2415 (1987).

\bibitem{Bogomolnyi90} E.~B. Bogomol'nyi, Sov. Phys. JETP {\bf 69}, 275
  (1989).

\bibitem{Goldberg91} J. Goldberg {\it et~al.}, Nonlinearity {\bf 4}, 1
  (1991); Goldberg and Schweizer, J. Phys. A {\bf 24}, 2785 (1991).

\bibitem{Taniguchi93d} N. Taniguchi and B.~L. Altshuler, Phys. Rev. Lett.
  {\bf 71}, 4031 (1993).

\bibitem{Prigodin94} A similar formula was used in V.~N. Prigodin {\it
    et.al.\/}, Phys. Rev. Lett.  {\bf 72}, 546 (1994) to study the time
  evolution of the density of quantum dots .

\end{references}

%
\input{epsf.tex}
\begin{figure}
%
\vspace{1cm}
\centerline{\epsfxsize=10cm \epsfbox{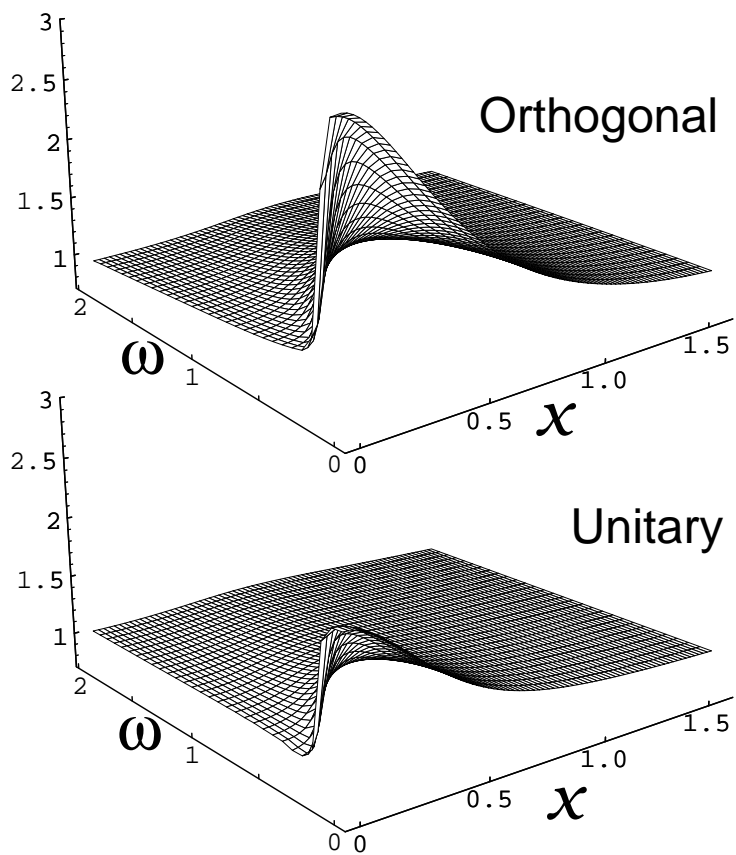}}
\vspace{2cm}
\caption{
Schematic plot of $1+s(\omega ,x)$ normalize by $1+k(\omega ,x)$
  for orthogonal (above) and unitary (below) cases.
}\label{fig:1}\end{figure}
\begin{figure}
\vspace{1cm}
\centerline{\epsfxsize=10cm \epsfbox{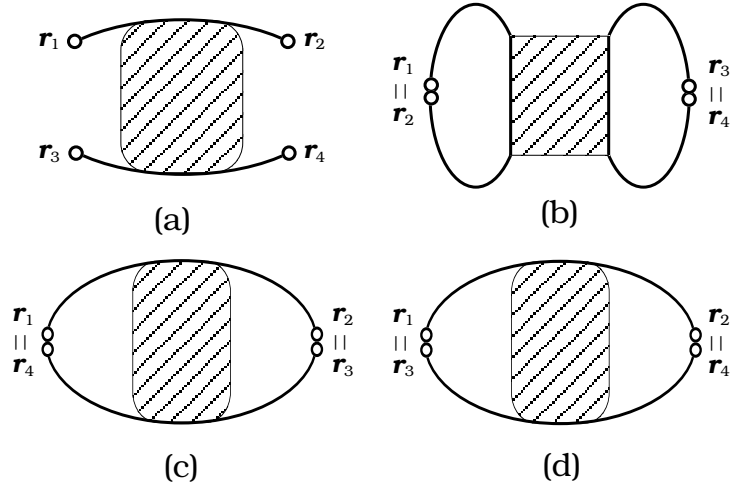}}
\vspace{1cm}
\caption{
  (a) Diagram representation of $R({\bf r}_1,{\bf r}_2,{\bf r}_3,{\bf
    r}_4)$.  (b) DOS fluctuation-type contribution.
  (c)~Cooperon-type contribution. (d)~Diffuson-type contribution.
  }\label{fig:2}\end{figure}




\end{document}